\documentstyle[twoside,fleqn,npb,epsfig,here,graphics]{article}
\newcommand{\pom}{{\rm I \hspace{-0.3ex} \rm P}}
\newcommand{\reg}{{\rm I \hspace{-0.3ex} \rm R}}
\newcommand{\der}{{\mathrm d}}

\newcommand{\AmS}{{\protect\the\textfont2
  A\kern-.1667em\lower.5ex\hbox{M}\kern-.125emS}}

\title{Diffractive Interactions: Experimental Summary}

\author{D.M. Jansen\address{Max--Planck--Institut f\"ur Kernphysik, Saupfercheckweg 1, 69117 Heidelberg, Germany}, 
M. Albrow\address{Fermilab, P.O. Box 500, Batavia, IL 60510--0500, U.S.A.} and
R. Brugnera\address{Dipartimento di Fisica dell' Universit\`a and INFN, Padova
, Italy}}

\begin{document}

\begin{abstract}
Experimental results on diffraction, which were 
presented at the $7$th International Workshop on Deep Inelastic
Scattering and QCD (DIS99), are summarized.
\end{abstract}

\maketitle

\section{Introduction}

During the two days of parallel sessions at DIS99, there were $19$
experimental talks about diffractive interactions 
on topics ranging from new measurements of ${\mathrm{F_2^{D(3)}}}$ 
at HERA to the observation of double--gap events at the Tevatron.
This paper summarizes the experimental results on diffraction
which were presented. The theoretical talks concerning diffractive
interactions are summarized in the following contribution.

\section{Inclusive Diffraction in DIS}

\begin{figure}[tb]
\epsfig{file=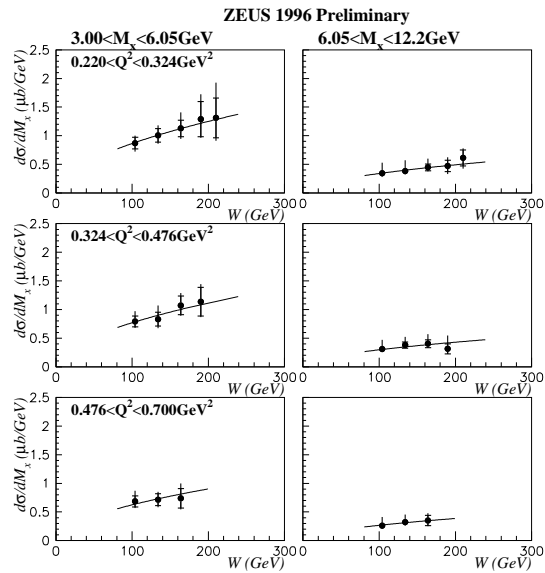,width=0.45\textwidth}
\vspace{-0.2in}
\caption{The differential cross section $\der \sigma/\der M_X$ as a function
of $W$. The solid line shows the fit result for
$\der \sigma/\der M_X \simeq W^{4 \overline{\alpha}_\pom-4}$.}
\label{Inuzuka1}
\end{figure}

M. Inuzuka presented the new ZEUS measurement of diffractive
cross sections at very low $Q^2$~\cite{Inuzuka}. Using 1996 data obtained
with their beam pipe calorimeter, ZEUS has measured $\der \sigma/\der M_X$ 
as a function of $W$
in the range $0.220 < Q^2 < 0.700$~GeV$^2$~(Figure~\ref{Inuzuka1}). 
Regge theory predicts that $\der \sigma/\der M_X \simeq
W^{4 \overline{\alpha}_\pom-4}$ and a fit to the data yields an effective 
($|t|$ averaged) pomeron intercept equal to: 
$\overline{\alpha}_\pom = 1.113 \pm 0.026$~(stat)~$^{+0.043}_{-0.062}$~(syst).
This value is approximately $1\sigma$ larger than the soft pomeron intercept
$\alpha_\pom(0) \simeq 1.09$ determined by Donnachie, Landshoff and Cudell~\cite{DL,Cudell}.
Assuming $\alpha_\pom^\prime = 0.25$~GeV$^{-2}$ and $|t| = 1/b$ with
$b=7.5$~GeV$^{-2}$, $\alpha_\pom(0) = \overline{\alpha}_\pom +0.033$.

\begin{figure}[tb]
\epsfig{file=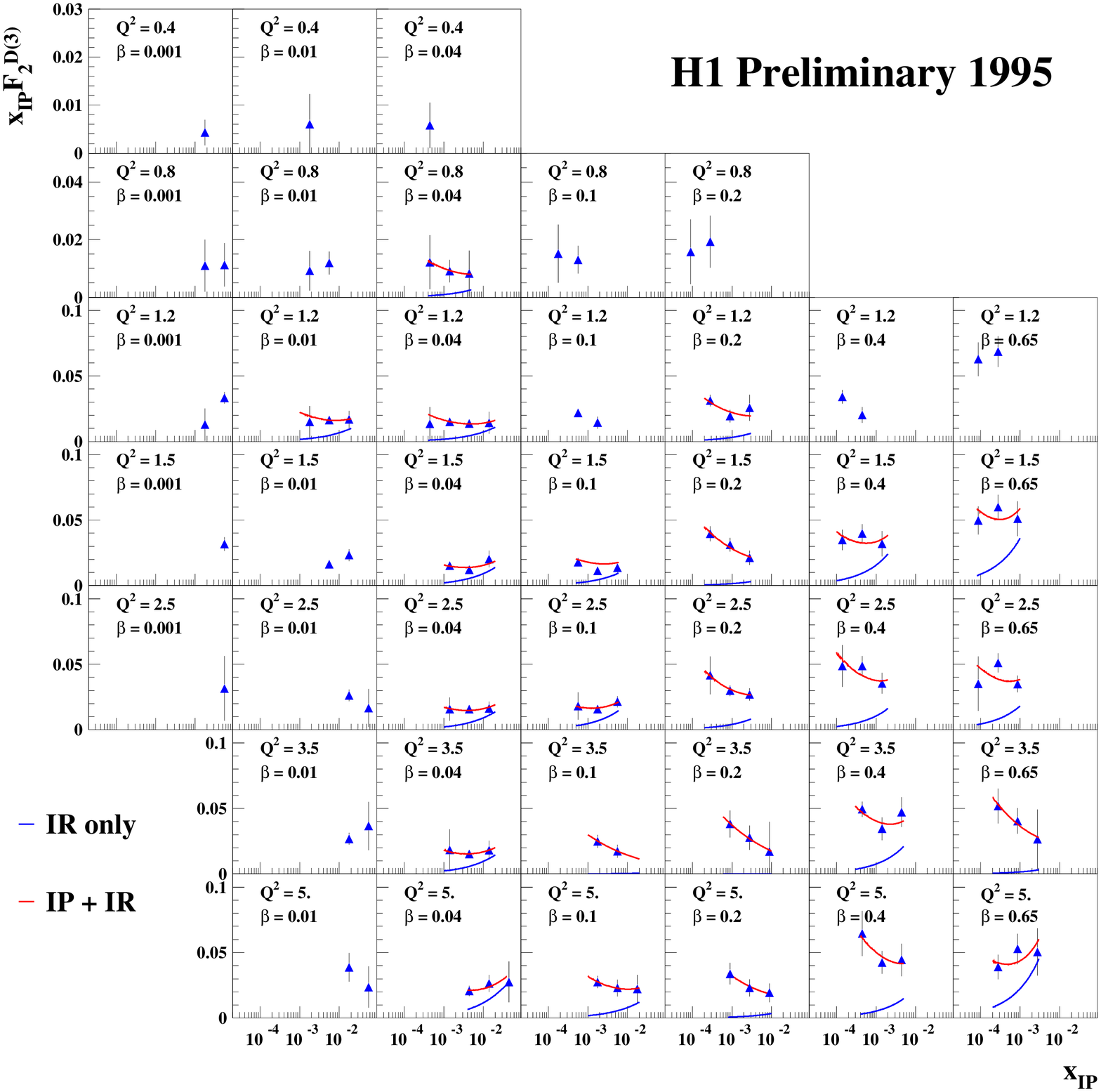,width=0.45\textwidth}
\vspace{-0.2in}
\caption{$x_{\pom} \cdot {\mathrm{F_2^{D(3)}}}$ shown as a function of 
$x_{\pom}$ in bins of $Q^2$ and $\beta$. The data are compared to a 
Regge based phenomenological fit.}
\label{Royon1}
\end{figure}

C. Royon presented the H1 preliminary measurement of 
${\mathrm{F_2^{D(3)}}}$ in the kinematic range $0.4 < Q^2 < 5$~GeV$^2$ and 
$0.001 < \beta < 0.65$~\cite{Royon}. The low $Q^2$ data were obtained during 
1995 when the interaction vertex was shifted by $70$ cm.
In Figure~\ref{Royon1}, the H1 measurement of 
$x_\pom \cdot {\mathrm{F_2^{D(3)}}}$
is compared to a phenomenological fit with diffractive ($\pom$) and 
sub--leading ($\reg$) exchange trajectories. The fitted pomeron intercept 
is consistent with the previous H1 measurement:
$\alpha_\pom(0) = 1.203 \pm 0.020$~(stat)~$\pm 0.013$~(syst)~$^{+0.030}_{-0.035}$~(model)~\cite{h1f2d3}. 
This value is significantly larger than the soft pomeron intercept.
The H1 collaboration also presented their 
measurement of ${\mathrm{F_2^{D(3)}}}$ in the high $Q^2$ range 
$200 < Q^2 < 800$~GeV$^2$. A QCD fit to the intermediate $Q^2$ data, 
with parton distributions for the pomeron and sub--leading reggeons which 
evolve according to the DGLAP~\cite{DGLAP} equations, gives a reasonable description of 
the high $Q^2$ data. 

\section{Hadronic Final State in Diffractive Interactions}

\begin{figure}[tb]
\vspace{-1.4in}
\includegraphics[bb=0 0 530 550, width=0.46\textwidth,clip=true]{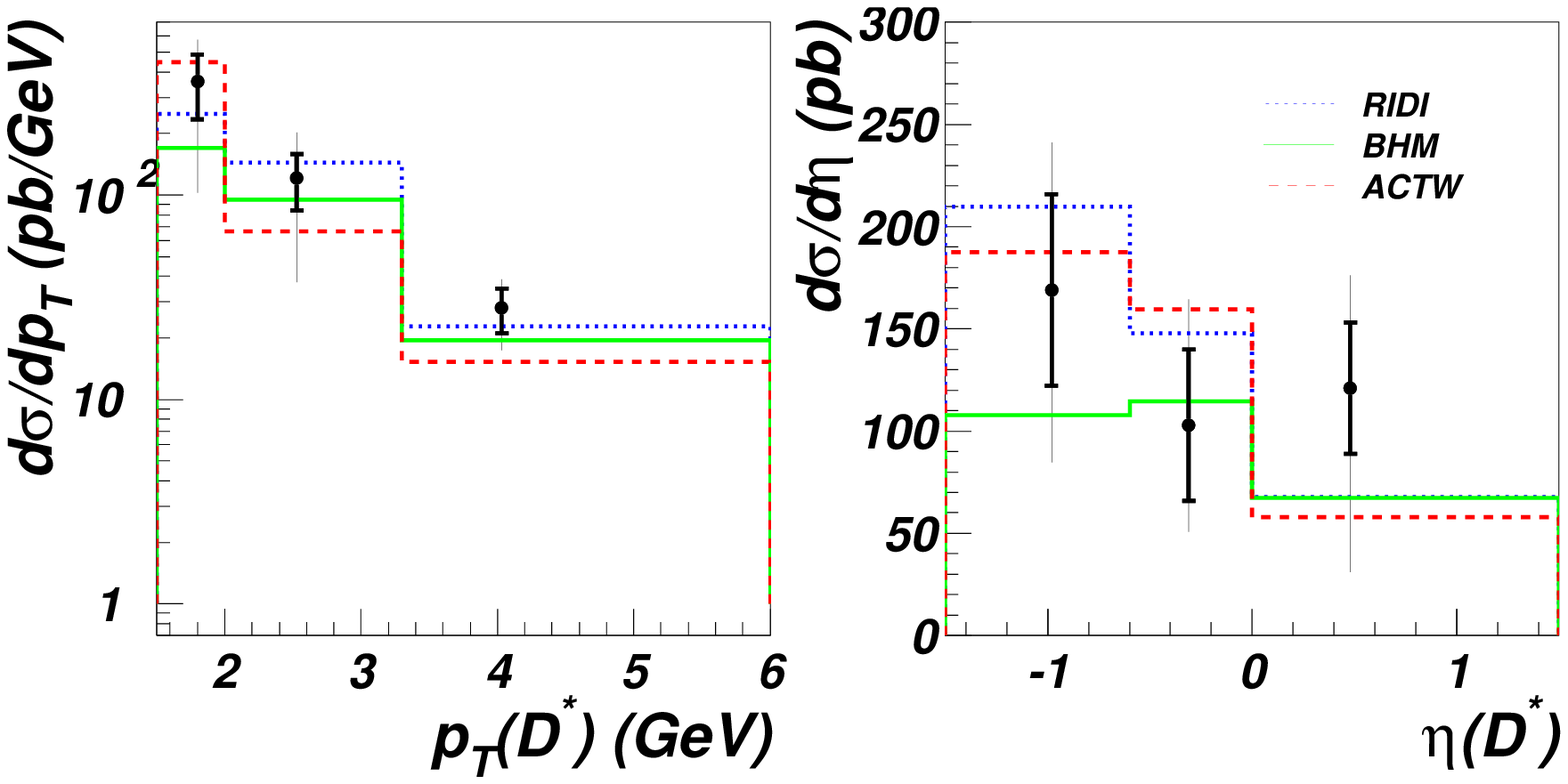}
\vspace{-0.4in}
\caption{Diffractive $D^{*\pm}$ cross sections 
$\der \sigma/\der p_T$ and $\der \sigma/\der \eta$
from the ZEUS collaboration compared to Monte Carlo predictions.}
\label{cole1}
\end{figure}

\begin{figure}[htb]
\mbox{
\epsfig{file=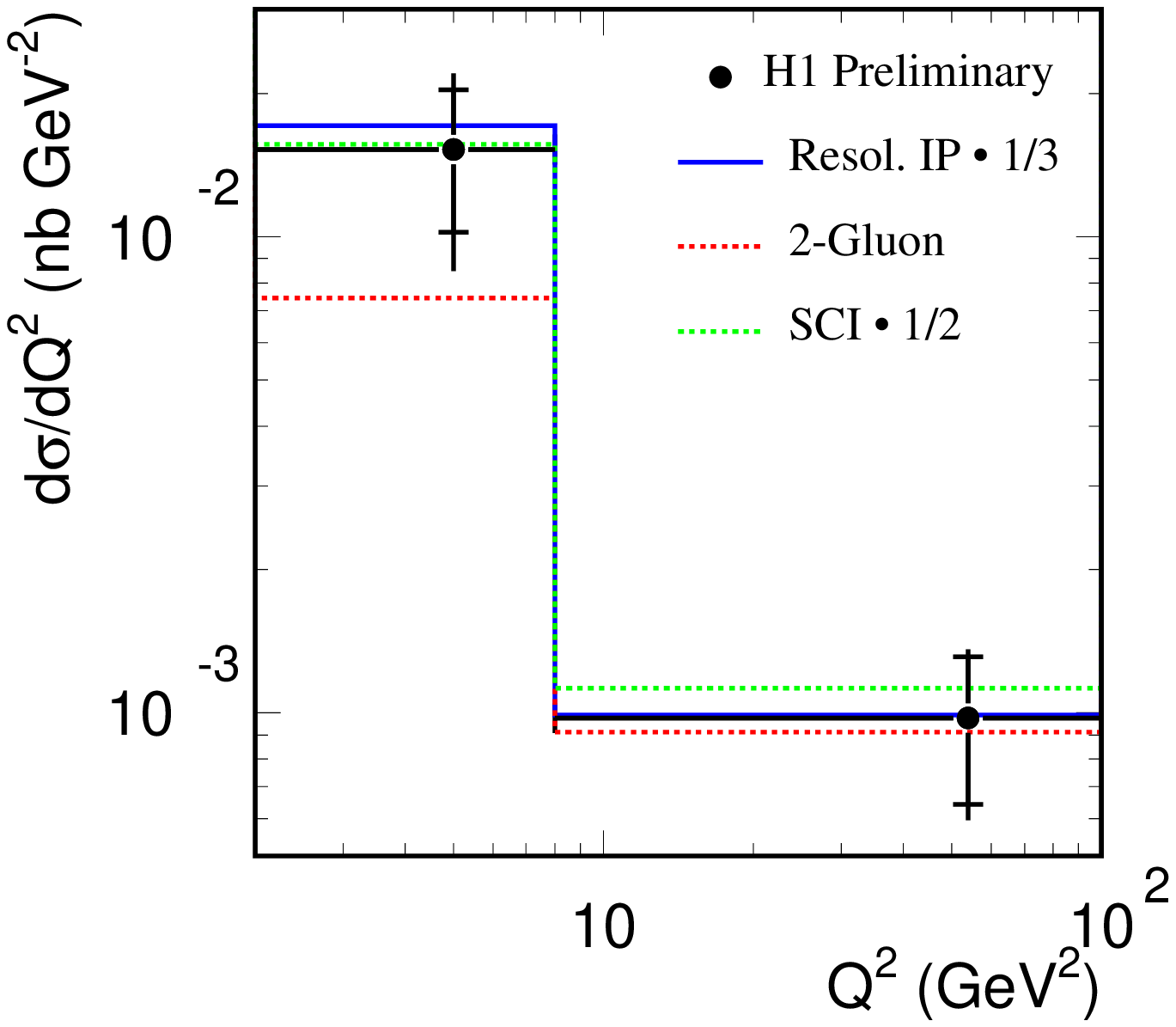,width=0.22\textwidth}
\epsfig{file=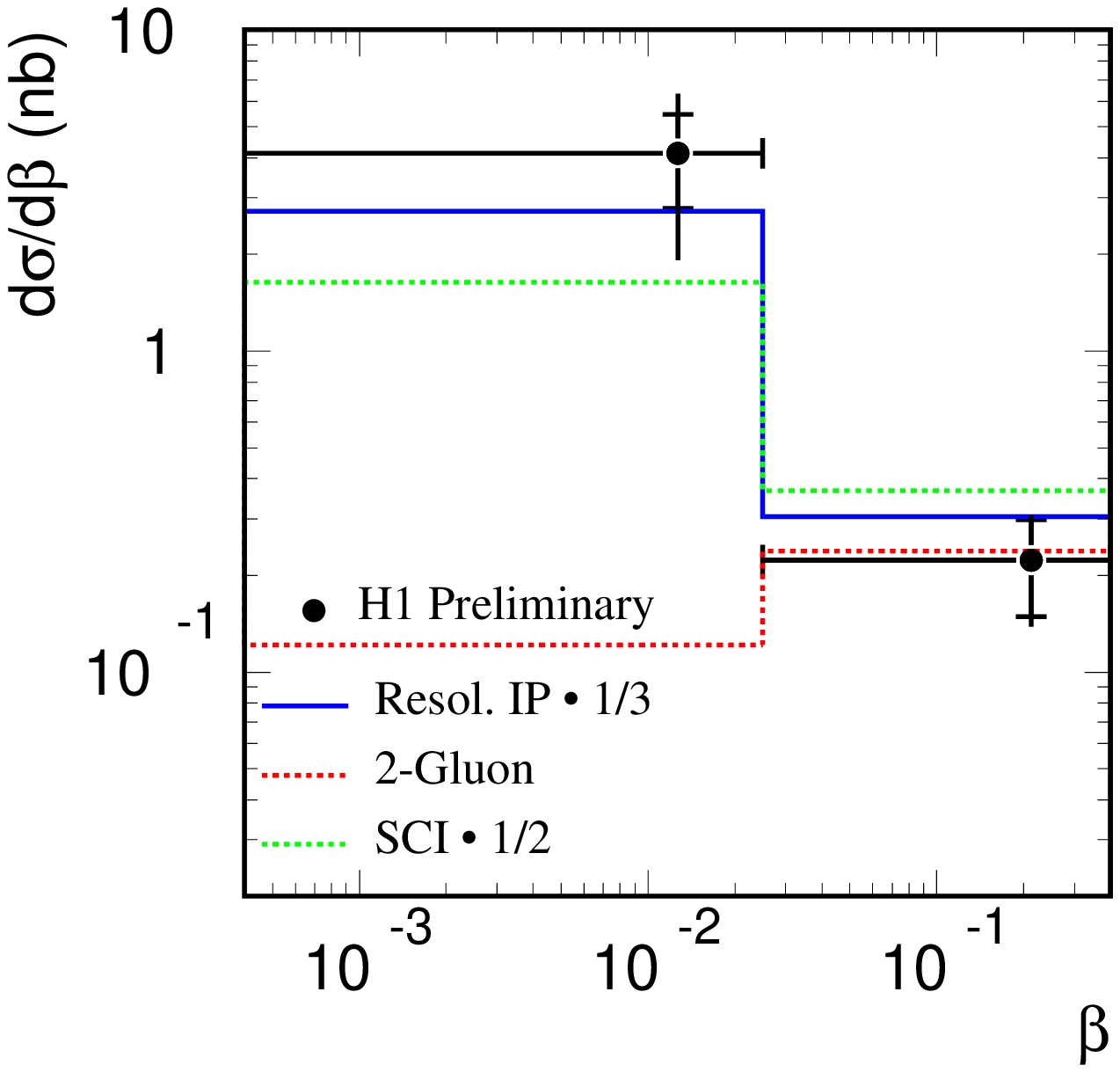,width=0.22\textwidth}
}
\vspace{-0.4in}
\caption{Diffractive $D^{*\pm}$ cross sections $\der \sigma/\der Q^2$
and $\der \sigma/\der \beta$ from the H1 collaboration compared to 
Monte Carlo predictions.}
\label{hengstmann1}
\end{figure}

The H1 collaboration has performed a NLO DGLAP analysis of their 
${\mathrm{F_2^{D(3)}}}$ measurement with $4.5 < Q^2 < 75$~GeV$^2$ and 
extracted diffractive parton distributions~\cite{h1f2d3}. These partons 
distributions when incorporated into Monte Carlo programs give a 
good description of the 
hadronic final state in hard diffractive processes. As an example of this, 
F.P. Schilling presented the H1 measurements of diffractive dijet 
production in $e p \rightarrow e X Y$ interactions with $M_Y < 1.6$~GeV 
and $|t| < 1$~GeV$^2$~\cite{Schilling}. 
The system $X$ contains two jets each with $p_T > 5$~GeV.
The POMPYT~\cite{pompyt} and RAPGAP~\cite{RAPGAP} Monte Carlo programs, 
with either a `flat' or a `peaked' gluon distribution, give good descriptions of the 
photoproduction and DIS data which were presented.
Predictions, calculated using diffractive 
parton distributions which consist solely of quarks at the starting scale, 
underestimate the measured cross sections by factors varying between 
$3$ and $6$. 

Diffractive $D^{*\pm}$ cross section measurements were
presented by S. Hengstmann and J. Cole for the H1 and ZEUS 
collaborations respectively~\cite{Hengstmann,Cole}. 
Both experiments use the 
$D^{*+} \rightarrow (D^0 \rightarrow K^- \pi^+) \pi^+ + $(c.c.)
decay mode whereas the ZEUS collaboration also uses the
$D^{*+} \rightarrow (D^0 \rightarrow K^- \pi^+ \pi^- \pi^+) \pi^+ +$(c.c.) 
decay mode. The cross section measurements for the two decay modes
from ZEUS are in excellent 
agreement when interpolated to the same kinematic region and are in good 
agreement with Monte Carlo calculations. The BHM~\cite{BHM} Monte Carlo prediction
shown in Figure~\ref{cole1} is based on soft colour interactions, whereas in the 
RIDI~\cite{RIDI} Monte Carlo diffractive charm production is proportional to the square 
of the gluon density in the proton. The ACTW~\cite{ACTW} Monte Carlo is a resolved 
pomeron model with diffractive parton distributions which evolve
according to the DGLAP~\cite{DGLAP} equations. The good agreement between data and model 
calculations shown in Figure~\ref{cole1}
is in contradiction with the results presented by S. Hengstmann. 
Figure~\ref{hengstmann1} shows that the H1 diffractive $D^{*\pm}$ cross 
sections are approximately a factor of $3$ smaller than the prediction from 
a resolved pomeron model (RAPGAP~\cite{RAPGAP}) and a factor of $2$ smaller than the prediction 
from a soft colour interactions model (AROMA~\cite{AROMA}).

\begin{figure}[tb]
\epsfig{file=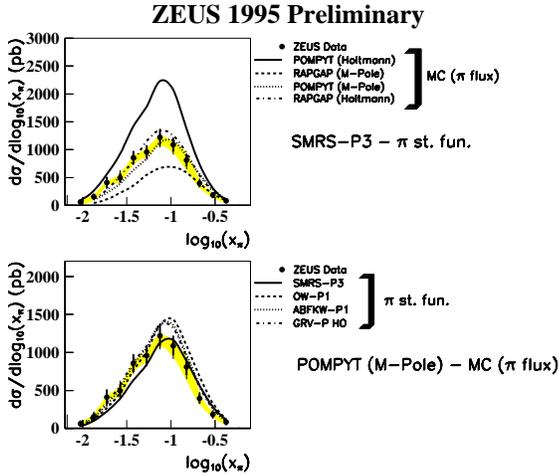,width=0.45\textwidth}
\vspace{-0.6in}
\caption{The photoproduction cross section $\der \sigma/\der {\rm log}(x_\pi)$
compared to Monte Carlo calculations with various pion flux factors (top)
and various pion parton distributions (bottom).}
\label{Khakzad1}
\end{figure}

M. Khakzad presented the new ZEUS measurement of dijet cross
sections associated with a leading neutron~\cite{Khakzad}. By requiring
a leading neutron with $E_n > 400$~GeV and $\theta_n < 0.8$~mrad,
$\pi^+$--exchange events are tagged. The fraction of the
pion's momentum participating in the production of the two jets can
be estimated using the final state variable
$x_\pi = \sum_{jets} E_T^{jets} e^{\eta^{jet}}/2E_\pi$. The 
ZEUS measurement of $\der \sigma/\der {\rm log}(x_\pi)$, presented in 
Figure~\ref{Khakzad1}, shows that in the kinematic region of the
measurement the $x_\pi$ distribution has only a mild sensitivity to 
the pion's structure function. The Monte Carlo calculations have a larger 
sensitivity to differences in the pion flux.

M. Kapichine presented the H1 measurements of semi--inclusive cross sections 
in the kinematic region $2 \le Q^2 \le 50$~{\rm GeV}$^2$,
$6\cdot 10^{-5} \le x \le 6\cdot 10^{-3}$ and baryon 
$p_T \le 200$~{\rm MeV}~\cite{Kapichine}. The semi--inclusive
cross sections are parameterized in terms of 
a leading baryon structure function, either ${\mathrm{F_2^{LP(3)}}}$ or 
${\mathrm{F_2^{LN(3)}}}$, for protons or neutrons respectively.
The leading baryon structure functions with $z > 0.7$ are reasonably well 
described by a Regge model of baryon production which considers the 
colour neutral exchanges of pions, pomerons and secondary reggeons. 
The semi--inclusive cross sections for leading neutrons can be described 
entirely by $\pi^+$ exchange whereas leading protons require $\pi^0$ and 
$f_2$ exchange contributions. The leading neutron data were used to estimate 
for the first time the structure function of the pion at small Bjorken--$x$.

\begin{figure}[tb]
\epsfig{file=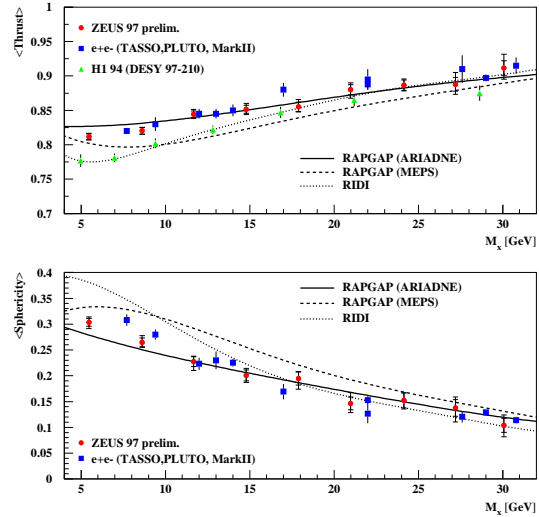,width=0.45\textwidth}
\vspace{-0.5in}
\caption{ Average thrust $<T>$ and sphericity $<S>$ as a function of 
$M_X$ for events tagged with a leading proton with $x_L > 0.95$.}
\label{Markun1}
\end{figure}

Results on leading baryon production from ZEUS were presented by
I. Gialas~\cite{Gialas}. The $x_L$ spectrum of leading protons
with $p_T^2 < 0.5$GeV$^2$ is well described by a Regge model of leading
baryon production whereas standard Monte Carlo programs, such as ARIADNE~\cite{ARIADNE}
and LEPTO~\cite{LEPTO}, fail to describe the data. The ratio of events with a leading 
baryon and all DIS events is independent of $x$ and $Q^2$ which  
supports the hypothesis of limiting fragmentation~\cite{Chou}.

P. Markun presented results on the hadronic final state in diffractive
DIS from ZEUS~\cite{Markun}. Figure~\ref{Markun1} shows the average
thrust and sphericity of diffractive events as a function of $M_X$
for events tagged with a leading proton with $x_L > 0.95$. 
The analysis was performed in the $\gamma^* \pom$ centre of mass system. 
The ZEUS measurements are compatible with the $\sqrt{s}$ dependence 
observed in $e^+ e^-$ events and are in fair agreement with the 
predictions from the RAPGAP~\cite{RAPGAP} Monte Carlo implemented with ARIADNE~\cite{ARIADNE} 
colour dipole fragmentation. The average thrust and sphericity
measurements are independent of $x_\pom$, $Q^2$ and $W$. Energy
flow measurements in the $\gamma^* \pom$ cms frame show that 
a two jet structure becomes increasingly pronounced as $M_X$ increases. 

\begin{figure}[tb]
\epsfig{file=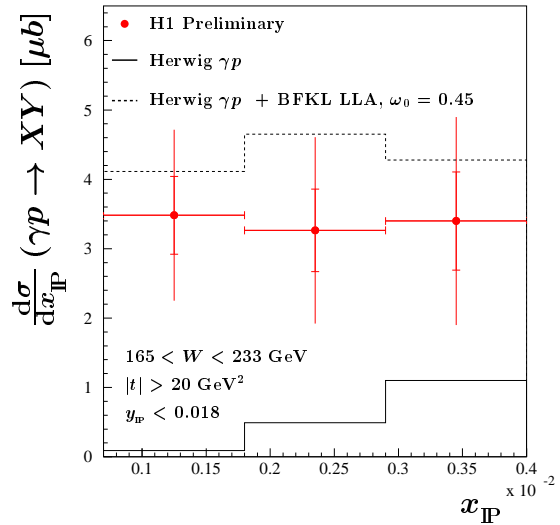,width=0.46\textwidth}
\vspace{-0.2in}
\caption{ The differential cross section 
$\der \sigma/\der x_\pom \, (\gamma \, p \rightarrow XY)$.}
\label{Cox1}
\end{figure}

B. Cox presented the H1 measurement of double diffractive dissociation at 
large $|t|$ in photoproduction~\cite{Cox}. The inclusive double
dissociative process $\gamma p \rightarrow X Y$ provides access to larger 
rapidity gaps than does the traditional gap--between--jets approach. This 
is advantageous since the BFKL cross section is expected to rise 
exponentially as a function of the rapidity separation~\cite{Tang}. 
In Figure~\ref{Cox1} the differential cross 
section $\der \sigma/\der x_\pom \, (\gamma \, p \rightarrow XY)$ is 
compared with the prediction from the HERWIG~\cite{HERWIG} Monte Carlo
for all non colour--singlet exchange processes. A significant excess above the 
expectation from the standard photoproduction model is observed. The dashed line
shows the HERWIG prediction with the LLA BFKL contribution added. Good
agreement is observed in both normalization and shape.

\begin{figure}[tb]
\epsfig{file=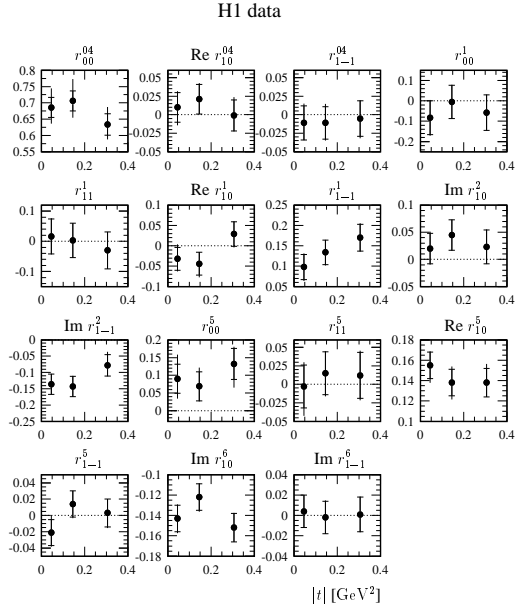,width=0.45\textwidth}
\vspace{-0.2in}
\caption{ The 15 spin density matrix elements for elastic 
electroproduction of $\rho$ mesons as a function of $|t|$. The
dashed lines indicate the expected null values in the case of SCHC.} 
\label{Clerbaux2}
\end{figure}

\section{Diffractive Vector Meson Production}

\begin{figure}[tb]
\epsfig{file=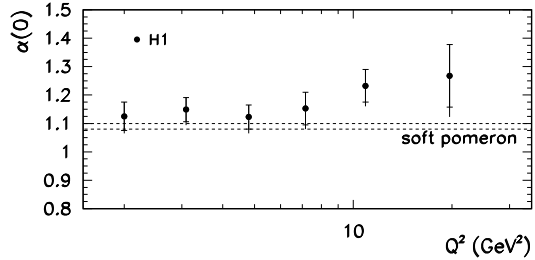,width=0.45\textwidth}
\vspace{-0.2in}
\caption{ The $Q^2$ dependence of $\alpha_\pom(0)$ for elastic $\rho$ 
electroproduction.}
\label{Clerbaux1}
\end{figure}

At DIS99 many new results on vector meson production were shown.
B. Clerbaux presented H1 results on elastic electroproduction of
$\rho$ mesons in the kinematic region $1 < Q^2 < 60$~GeV$^2$ and
$30 < W < 140$~GeV~\cite{Clerbaux}. Results on the shape of the
$(\pi \pi)$ mass distribution were presented which indicate
significant skewing at low $Q^2$, which gets smaller with increasing
$Q^2$. No significant $W$ or $|t|$ dependence of the skewing is
observed. Measurements of the 15 elements of the $\rho$ spin density matrix 
were also presented as a function of $Q^2$, $W$ and $|t|$
(see Figure~\ref{Clerbaux2} for example).
Except for a small but significant deviation from zero of the 
$r_{00}^{5}$ matrix element, 
s--channel helicity conservation (SCHC) is found to be a good approximation.
The dominant helicity flip amplitude $T_{01}$ is $(8 \pm 3)\%$ of the
non--flip amplitudes. The $W$ dependence of the measured  
$\gamma^* p \rightarrow \rho p$ cross sections, for six fixed values of 
$Q^2$ (Figure~\ref{Clerbaux1}),
suggests that the effective trajectory governing $\rho$ 
electroproduction is larger than the soft pomeron intercept determined by 
Donnachie, Landshoff and Cudell~\cite{DL,Cudell}. 

\begin{figure}[tb]
\epsfig{file=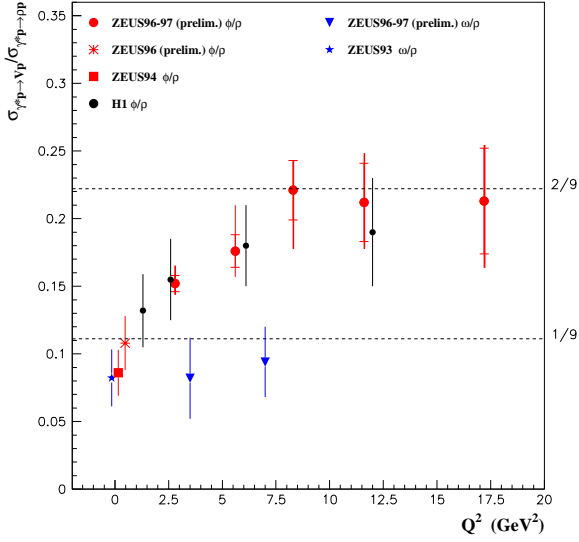,width=0.46\textwidth}
\vspace{-0.2in}
\caption{ The $Q^2$ dependence of vector meson cross section ratios.}
\label{Savin1}
\end{figure}

Preliminary results from H1 on proton dissociative $\rho$ meson
production were presented by A. Droutskoi~\cite{Droutskoi}. 
Proton dissociative events are tagged by requiring activity in either 
the forward part $(\eta > 2.7)$ of the LAr calorimeter, the forward muon 
detector or the proton remnant detector. An indication is observed for
an increase in the ratio of proton dissociative to elastic $\rho$ meson
cross sections in the region $1.5 < Q^2 < 3$~GeV$^2$, in contrast with 
the approximately flat behavior of the ratio in the region $Q^2 > 3$~GeV$^2$.
Results were also presented on the angular distributions characterizing 
the $\rho$ meson production and decay.

New results on exclusive $\omega$ meson production from ZEUS were presented by 
A. Savin~\cite{Savin}. In the kinematic range $ 40 < W < 120$~GeV and 
$3 < Q^2 < 30$~GeV$^2$, 
$\sigma (\gamma^* p \rightarrow \omega p) \simeq W^{0.7}$ and 
$\sigma (\gamma^* p \rightarrow \omega p) \simeq 1/(Q^2 + M_\omega^2)^2$.
These dependencies are consistent with those found for the $\rho$. 
The ratio of $\rho:\omega:\phi$ production, which is shown in 
Figure~\ref{Savin1}, is in good agreement at large $Q^2$
with the SU(3) prediction $9:1:2$. Exclusive cross sections, 
for the production of $\rho$, $\phi$ and $\omega$ vector mesons, 
are found to be proportional to $W^\delta$ where $\delta$ increases with $Q^2$. 
Results were also presented which show that SCHC is violated for
exclusive $\rho$ meson production in the low $Q^2$ range $0.25 < Q^2 < 0.85$~GeV$^2$. 

\begin{figure}[tb]
\epsfig{file=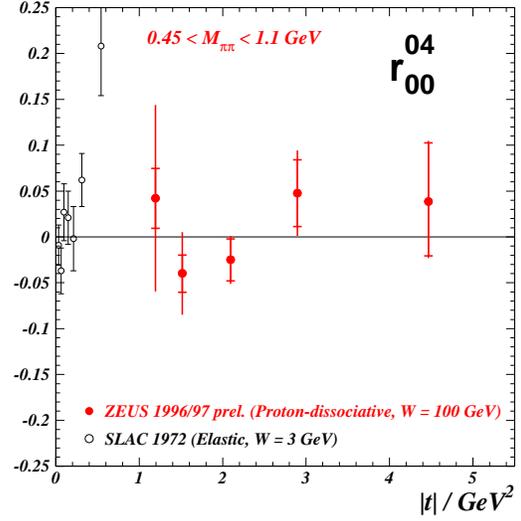,width=0.45\textwidth}
\vspace{-0.2in}
\caption{ The spin density matrix element $r_{00}^{04}$ as a function
of $|t|$ for $\rho^0$ photoproduction.}
\label{Crittenden1}
\end{figure}

J. Crittenden presented new results from the ZEUS collaboration on
$\rho^0$ photoproduction at high momentum transfer $|t|$~\cite{Crittenden}.
Measurements of $r_{00}^{04}$ (Figure~\ref{Crittenden1}),
in the range $1 < |t| < 9$~GeV$^2$, were presented which show that the $\rho^0$ does not 
become dominantly longitudinally polarized at large values of $|t|$~\cite{Ginzburg,Diehl}. 
Evidence for a non--zero value of the double--flip amplitude 
$r_{1-1}^{04}$ at large $|t|$ was also presented. 

P. Merkel presented new H1 results on the elastic production of $J/\psi$ and
$\psi(2S)$ mesons in the kinematic region $2 < Q^2 < 80$~GeV$^2$ and
$25 < W < 160$~GeV~\cite{Merkel}. The dependence of the cross section
$\sigma(\gamma^* p \rightarrow J/\psi p)$ on $W$ is proportional to $W^\delta$,
with $\delta \simeq 1$ which has also been observed in photoproduction.
The $Q^2$ dependence is proportional to $1/(Q^2 + m_{J/\psi}^2)^n$
with $n=2.38 \pm 0.11$. The first evidence from HERA for the quasi--elastic production
of $\psi(2S)$ in DIS was also reported. The ratio of cross sections for $\psi(2S)$ and
$J/\psi$ production increases as a function of $Q^2$.

Results on exclusive $\rho^0$ electroproduction from the HERMES collaboration were 
presented by A. Borissov~\cite{Borissov}. Using ${\rm ^1H}$, ${\rm ^2H}$, ${\rm ^3He}$ 
and ${\rm ^{14}N}$
targets, the ratio of cross sections per nucleon $\sigma_A/(A \sigma_H)$,
known as the nuclear transparency, was found to decrease with increasing coherence length 
of quark--antiquark fluctuations of the virtual photon. An unperturbed virtual
state with mass $M_{q\overline{q}}$ can travel a coherence length distance
$l_c = 2\nu/(Q^2+M_{q\overline{q}}^2)$ in the laboratory frame during its lifetime. 
The data presented showed clear evidence for the interaction of the quark--antiquark 
fluctuations with the nuclear medium.

\section{Diffraction at the Tevatron and LEP Colliders}

\begin{figure}[tb]
\includegraphics[bb=55 30 540 650, width=0.46\textwidth,clip=true]{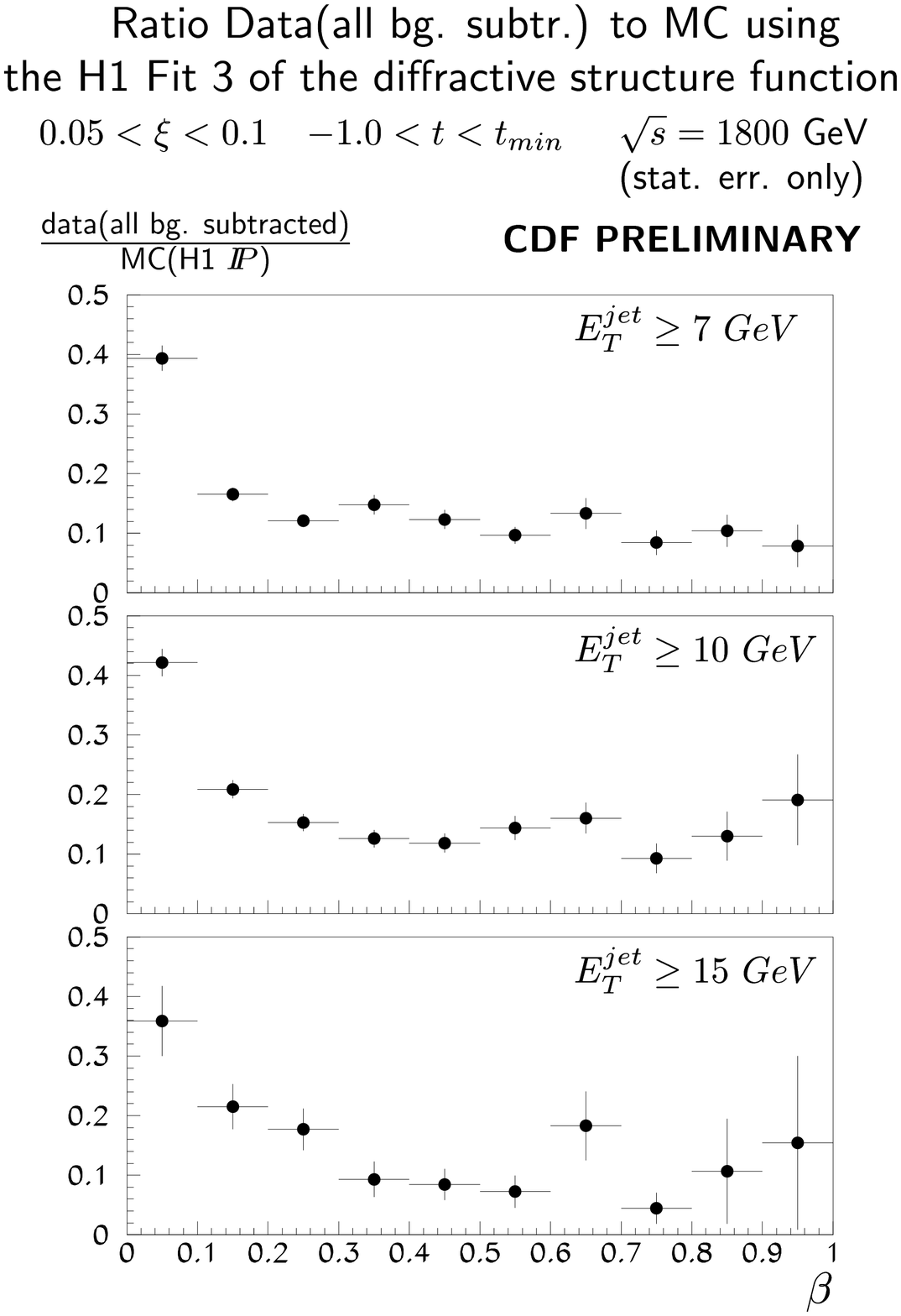}
\vspace{-0.2in}
\caption{Ratio of CDF diffractive dijet data and Monte Carlo predictions as a 
function of $\beta$. The Monte Carlo predictions were calculated using the 
H1 `peaked' gluon distribution for the pomeron.}
\label{borras1}
\end{figure}

Results on hard diffraction from CDF were presented by K. Borras~\cite{Borras}.
Using a sample of diffractive dijet events, with a recoil beam particle tagged with 
Roman Pot detectors, CDF has extracted the momentum fraction of the interacting
parton in the pomeron using the formula $\beta = (E_T^{jet1} e^{-\eta^{jet1}}
+ E_T^{jet2} e^{-\eta^{jet2}})/ 2 \xi P_{beam}$ where $\xi$ is the momentum fraction 
of the beam particle taken by the pomeron. After subtracting background contributions,
such as non--diffractive dijet production with an accidental hit in the Roman Pot detectors
and the contribution due to meson exchange, the data were compared to Monte Carlo 
simulations assuming various pomeron parton distributions and pomeron flux parametrizations. 
Figure~\ref{borras1} shows for three jet energy thresholds the ratio of the 
CDF diffractive dijet data and Monte Carlo predictions as a function of $\beta$. The
Monte Carlo predictions were calculated with the H1 `peaked' gluon distribution~\cite{h1f2d3}
for the pomeron. The ratios are flat for $\beta > 0.2$  
and approximately equal to 0.15. This result implies that the $\beta$ distributions agree well 
with the shape of the H1 pomeron structure function but that there is a discrepancy 
between the data and the normalization of the standard pomeron flux. For $\beta > 0.2$, 
the shape and rate of the $\beta$ distributions agree well with Monte Carlo predictions 
calculated with Goulianos's renormalized flux~\cite{Goulianos} and a flat gluon 
distribution, although an enhancement still exists in the $\beta < 0.2$ region.

\begin{figure}[tb]
\vspace{-1.4in}
\includegraphics[bb=0 0 530 550, width=0.46\textwidth,clip=true]{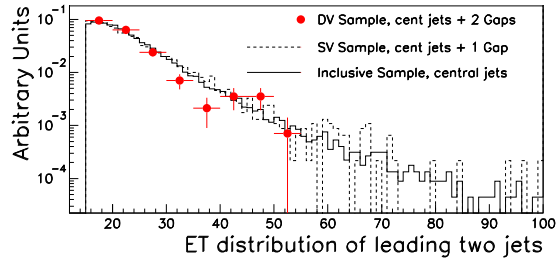}
\vspace{-0.2in}
\caption{The $E_T$ distributions of the leading two jets in double--gap, single diffractive
and inclusive interactions from the D0 collaboration.}
\label{mauritz1}
\end{figure}

\begin{figure}[tb]
\epsfig{file=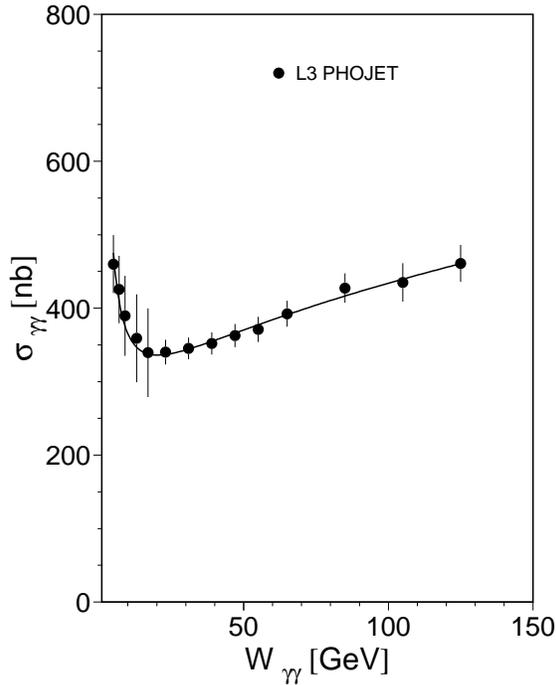,width=0.46\textwidth}
\vspace{-0.2in}
\caption{The L3 measurement of  
$\sigma_{\gamma \gamma}$ as a function of $W_{\gamma \gamma}$.
The data were corrected for acceptance and efficiency effects using the
PHOJET~\cite{PHOJET} Monte Carlo. The solid line shows the result of a 
Donnachie and Landshoff type fit for the total cross section.}
\label{vogt1}
\end{figure}

K. Mauritz presented results on hard diffraction from D0~\cite{Mauritz}. The fraction
of single diffractive dijet events at $\sqrt{s} = 1800$~GeV and 630~GeV were compared
to Monte Carlo predictions. Predictions calculated with hard or flat gluon distributions
give rates which are higher than observed in the data. For example, the fraction
of 1800 FWD Jet events observed in the data is $(0.64 \pm 0.05)$\% compared to the
hard and soft gluon predictions equal to $(2.1 \pm 0.3)$\% and $(1.6 \pm 0.3)$\%
respectively.
A comparison was also presented 
between the $E_T$ spectra of the leading two jets in double--gap events and the $E_T$
spectra observed in single diffractive and inclusive interactions. In spite of the 
decreasing effective centre of mass energies 
$(\sqrt{s}_{DG} < \sqrt{s}_{SD} < \sqrt{s}_{INC})$, Figure~\ref{mauritz1} shows that 
the $E_T$ distributions are similar in shape which suggests a hard structure for the 
pomeron. A study of event characteristics shows that diffractive events are quieter, and 
contain thinner jets, than do non--diffractive events. The same conclusions
have been reached by CDF.

H. Vogt presented L3 results on hadron production in $\gamma \gamma$
collisions at LEP~\cite{Vogt}. Cross sections measurements 
with quasi--real photons (anti--tagged events where the scattered
electrons are not detected) and with virtual photons (where both
scattered electrons are detected in small angle calorimeters)
were presented. Figure~\ref{vogt1} shows the measurement of 
$\sigma(\gamma \gamma \rightarrow {\rm hadrons})$ as a function of $W_{\gamma \gamma}$
compared to a Donnachie and Landshoff type fit $(\sigma_{\rm tot} = A s^\epsilon + B s^{-\eta})$
for the total cross section.
The fit result for the pomeron dependence,
$\epsilon = 0.158 \pm 0.006$~(stat)~$\pm 0.028$~(syst), is 
a factor of two higher than the soft pomeron intercept. Fits to the measured  
$\sigma(\gamma^* \gamma^* \rightarrow {\rm hadrons})$ cross sections at
$\sqrt{s} = 91$, 183 and 189~GeV yield $\epsilon$ values equal to $0.28 \pm 0.05$,
$0.40 \pm 0.07$ and $0.29 \pm 0.03$ respectively.  These values are not in agreement with a 
leading order BFKL model with $\epsilon \simeq 0.53$~\cite{Brodsky}.

\end{document}